\documentclass{llncs}
\usepackage{amsmath}
\usepackage{amssymb}

\usepackage[ruled,vlined]{algorithm2e}

\begin{document}

%\date{}
%\renewcommand{\thefootnote}{\fnsymbol{footnote}}
%\addtocounter{footnote}{0}

\title{Linear Time Algorithms for Finding \\
a Dominating Set of Fixed Size \\
in Degenerated Graphs}
\titlerunning{Linear Time Algorithms for Finding a Dominating Set of Fixed Size}

\author{Noga Alon \inst{1} \and Shai Gutner \inst{2}}
\institute{Schools of Mathematics and Computer Science, Tel-Aviv
University, Tel-Aviv, 69978, Israel.
\thanks{Research supported in part by a grant from the Israel
Science Foundation, and by the Hermann Minkowski Minerva Center
for Geometry at Tel Aviv University.}
\email{noga@math.tau.ac.il.}
\and School of Computer Science, Tel-Aviv University, Tel-Aviv,
69978, Israel.
\thanks{This paper forms part of a Ph.D. thesis
written by the author under the supervision of
Prof. N. Alon and Prof. Y. Azar in Tel Aviv University.}
\email{gutner@tau.ac.il.}}

\maketitle

\begin{abstract}
There is substantial literature dealing with fixed parameter
algorithms for the dominating set problem on various families of
graphs. In this paper, we give a $k^{O(dk)} n$ time algorithm for
finding a dominating set of size at most $k$ in a $d$-degenerated
graph with $n$ vertices. This proves that the dominating set problem is
fixed-parameter tractable for degenerated graphs. For graphs that
do not contain $K_h$ as a topological minor, we give an improved
algorithm for the problem with running time $(O(h))^{hk} n$. For
graphs which are $K_h$-minor-free, the running time is further
reduced to $(O(\log h))^{hk/2} n$. Fixed-parameter tractable
algorithms that are linear in the number of vertices of the graph
were previously known only for planar graphs.

For the families of graphs discussed above, the problem of finding
an induced cycle of a given length is also addressed. 
For every fixed $H$ and $k$, we
show that if an $H$-minor-free graph $G$ with $n$ vertices
contains an induced cycle of size $k$, then such a cycle can be
found in $O(n)$ expected time as well as in $O(n \log n)$ worst-case
time. Some results are stated concerning the (im)possibility of
establishing linear time algorithms for the more general family of
degenerated graphs.


\end{abstract}

\section{Introduction}\label{sec:intro}

This paper deals with fixed-parameter algorithms for degenerated
graphs. The degeneracy $d(G)$ of an undirected graph $G=(V,E)$ is
the smallest number $d$ for which there exists an acyclic
orientation of $G$ in which all the outdegrees are at most $d$.
Many interesting families of graphs are degenerated (have bounded
degeneracy). For example, graphs embeddable on some fixed surface,
degree-bounded graphs, graphs of bounded tree-width, and
non-trivial minor-closed families of graphs.

There is an extensive literature dealing with fixed-parameter
algorithms for the dominating set problem on various families of
graphs. Our main result is a linear time algorithm for finding a dominating
set of fixed size in degenerated graphs. This is the most general
class of graphs for which fixed-parameter tractability for this problem has been
established. To the best of our knowledge, linear time algorithms
for the dominating set problem were previously known only for planar
graphs. Our algorithms both generalize and simplify the classical
bounded search tree algorithms for this problem (see, e.g.,
\cite{journals/jcss/AlberFFFNRS05,journals/jal/EllisFF04}).

The problem of finding induced cycles in degenerated graphs has
been studied by Cai, Chan and Chan \cite{conf/iwpec/CaiCC06}.  Our second result in
this paper is a randomized algorithm for finding an induced
cycle of fixed size in graphs with an excluded minor. The
algorithm's expected running time is linear, and its
derandomization is done in an efficient way, answering an open
question from \cite{conf/iwpec/CaiCC06}. The problem of finding
induced cycles in degenerated graphs is also addressed.

\textbf{The Dominating Set Problem.} The dominating set problem on
general graphs is known to be $W[2]$-complete \cite{MR1656112}.
This means that most likely there is no $f(k) \cdot n^c$-algorithm
for finding a dominating set of size at most $k$ in a graph of
size $n$ for any computable function $f: \bbbn \to
\bbbn$ and constant $c$. This suggests the exploration of
specific families of graphs for which this problem is
fixed-parameter tractable. For a general introduction to the field
of parameterized complexity, the reader is referred to
\cite{MR1656112} and \cite{MR2238686}.

The method of bounded search trees has been used to give an $O(8^k n)$ time
algorithm for the dominating set problem in planar graphs
\cite{journals/jcss/AlberFFFNRS05} and an $O((4g+40)^k n^2)$ time
algorithm for the problem in graphs of bounded genus $g \geq 1$
\cite{journals/jal/EllisFF04}. The algorithms for planar graph
were improved to $O(4^{6 \sqrt{34k}} n)$ \cite{Alber:2002:FPA},
then to $O(2^{27 \sqrt{k}} n)$ \cite{conf/mfcs/KanjP02}, and finally to
$O(2^{15.13 \sqrt{k}} k + n^3 + k^4)$ \cite{conf/soda/FominT03}.
Fixed-parameter algorithms are now known also for map graphs
\cite{Demaine:2005:FPA} and for constant powers of $H$-minor-free
graphs \cite{Demaine:2005:SPA}. The running time given in
\cite{Demaine:2005:SPA} for finding a dominating set of size $k$
in an $H$-minor-free graph $G$ with $n$ vertices is
$2^{O(\sqrt{k})} n^c$, where $c$ is a constant depending only on
$H$. To summarize these results, fixed-parameter tractable
algorithms for the dominating set problem were known for fixed
powers of $H$-minor-free graphs and for map graphs. Linear time
algorithms were established only for planar graphs.

\textbf{Finding Paths and Cycles.} The foundations for the
algorithms for finding cycles, presented in this paper, have been
laid in \cite{JACM::AlonYZ1995}, where the authors introduce the color-coding
technique. Two main randomized algorithms
are presented there, as follows. A simple directed or undirected
path of length $k-1$ in a graph $G=(V,E)$ that contains such a
path can be found in $2^{O(k)}|E|$ expected time in the directed
case and in $2^{O(k)}|V|$ expected time in the undirected case. A
simple directed or undirected cycle of size $k$ in a graph
$G=(V,E)$ that contains such a cycle can be found in either
$2^{O(k)} |V||E|$ or $2^{O(k)}|V|^\omega$ expected time, where
$\omega < 2.376$ is the exponent of matrix multiplication. 
These algorithms can be derandomized at a cost of an extra $\log |V|$
factor. As for the case of even cycles, it is shown in
\cite{Yuster:1997:FEC} that for every fixed $k \geq 2$, there is
an $O(|V|^2)$ algorithm for finding a simple cycle of size $2k$ in
an undirected graph (that contains such a cycle). Improved algorithms for detecting given
length cycles have been presented in \cite{ALGOR::AlonYZ1997} and
\cite{conf/soda/YusterZ04}. The authors of 
\cite{ALGOR::AlonYZ1997} describe
fast algorithms for finding short cycles in
$d$-degenerated graphs. In particular, $C_3$'s and $C_4$'s can be
found in $O(|E| \cdot d(G))$ time and $C_5$'s in $O(|E| \cdot
d(G)^2)$ time.

\textbf{Finding Induced Paths and Cycles.} Cai, Chan and Chan have
recently introduced a new interesting technique they call \textit{random
separation} for solving fixed-cardinality optimization problems on
graphs \cite{conf/iwpec/CaiCC06}. They combine this technique together with color-coding to
give the following algorithms for finding an induced graph within
a large graph. For fixed constants $k$ and $d$, if a
$d$-degenerated graph $G$ with $n$ vertices contains some fixed
induced tree $T$ on $k$ vertices, then it can be found in $O(n)$
expected time and $O(n \log^2 n)$ worst-case time. If such a graph
$G$ contains an induced $k$-cycle, then it can be found in
$O(n^2)$ expected time and $O(n^2 \log^2 n)$ worst-case time. Two
open problems are raised by the authors of the paper. First, they
ask whether the $\log^2 n$ factor incurred in the derandomization
can be reduced to $\log n$. A second question is whether there is
an $O(n)$ expected time algorithm for finding an induced $k$-cycle
in a $d$-degenerated graph with $n$ vertices. In this paper, we
show that when combining the techniques of random separation and
color-coding, an improved derandomization with a loss of only $\log
n$ is indeed possible. An $O(n)$ expected time algorithm finding
an induced $k$-cycle in graphs with an excluded minor is
presented. We give evidence that establishing such an algorithm
even for 2-degenerated graphs has far-reaching consequences.

\textbf{Our Results.} The main result of the paper is that the
dominating set problem is fixed-parameter tractable for
degenerated graphs. The running time is $k^{O(dk)} n$ for finding
a dominating set of size $k$ in a $d$-degenerated graph with $n$
vertices. The algorithm is linear in the number of vertices of the
graph, and we further improve the dependence on $k$ for the
following specific families of degenerated graphs. For graphs that
do not contain $K_h$ as a topological minor, an improved algorithm
for the problem with running time $(O(h))^{hk} n$ is established.
For graphs which are $K_h$-minor-free, the running time obtained
is $(O(\log h))^{hk/2} n$. We show that all the algorithms can be
generalized to the weighted case in the following sense. A dominating set of
size at most $k$ having minimum weight can be found within the
same time bounds.

We address two open questions raised by Cai, Chan and Chan in
\cite{conf/iwpec/CaiCC06} concerning linear time algorithms for
finding an induced cycle in degenerated graphs. An $O(n)$ expected
time algorithm for finding an induced $k$-cycle in graphs with an
excluded minor is presented. The derandomization performed in
\cite{conf/iwpec/CaiCC06} is improved and we get a deterministic
$O(n \log n)$ time algorithm for the problem. As for finding
induced cycles in degenerated graphs, we show a deterministic
$O(n)$ time algorithm for finding cycles of size at most $5$, and
also explain why this is unlikely to be possible to achieve for longer
cycles.

\textbf{Techniques.} We generalize the known search tree
algorithms for the dominating set problem. This is enabled by
proving some combinatorial lemmas, which are interesting in their
own right. For degenerated graphs, we bound the number of vertices
that dominate many elements of a given set, whereas for graphs with
an excluded minor, our interest is in vertices that still need to
be dominated and have a small degree.

The algorithm for finding an induced cycle in non-trivial
minor-closed families is based on 
random separation and color-coding. Its 
derandomization is performed using known explicit constructions of
families of (generalized) perfect hash functions.

\section{Preliminaries}

The paper deals with undirected and simple graphs, unless stated
otherwise. Generally speaking, we will follow the notation used in
\cite{MR0411988} and \cite{MR2159259}. For an undirected graph
$G=(V,E)$ and a vertex $v \in V$, $N(v)$ denotes the set of
all vertices adjacent to $v$ (not including $v$ itself). We say
that $v$ \textit{dominates} the vertices of $N(v) \cup \{v\}$. The graph
obtained from $G$ by deleting $v$ is denoted $G-v$. The subgraph
of $G$ induced by some set $V' \subseteq V$ is denoted by $G[V']$.

A graph $G$ is \textit{$d$-degenerated} if every induced subgraph
of $G$ has a vertex of degree at most $d$. It is easy and known that every
$d$-degenerated graph $G=(V,E)$ admits an acyclic orientation such
that the outdegree of each vertex is at most $d$. Such an
orientation can be found in $O(|E|)$ time. A $d$-degenerated graph
with $n$ vertices has less than $dn$ edges and therefore its average
degree is less than $2d$.

For a directed graph $D=(V,A)$ and a vertex $v \in V$, the set of
out-neighbors of $v$ is denoted by $N^+(v)$. For a set $V'
\subseteq V$, the notation $N^+(V')$ stands for the set of all
vertices that are out-neighbors of at least one vertex of $V'$.
For a directed graph $D=(V,A)$ and a vertex $v \in V$, we define
$N_1^+(v)=N^+(v)$ and $N_i^+(v)=N^+(N_{i-1}^+(v))$ for $i \geq 2$.

An edge is said to be \textit{subdivided} when it is deleted and
replaced by a path of length two connecting its ends, the internal
vertex of this path being a new vertex. A \textit{subdivision} of
a graph $G$ is a graph that can be obtained from $G$ by a sequence
of edge subdivisions. If a subdivision of a graph $H$ is the
subgraph of another graph $G$, then $H$ is a \textit{topological
minor} of $G$. A graph $H$ is called a \textit{minor} of a graph
$G$ if is can be obtained from a subgraph of $G$ by a series of
edge contractions.

In the parameterized dominating set problem, we are given an
undirected graph $G=(V,E)$, a parameter $k$, and need to find a
set of at most $k$ vertices that dominate all the other vertices.
Following the terminology of \cite{journals/jcss/AlberFFFNRS05},
the following generalization of the problem is considered. The
input is a \textit{black and white graph}, which simply means that
the vertex set $V$ of the graph $G$ has been partitioned into two
disjoint sets $B$ and $W$ of black and white vertices,
respectively, i.e., $V=B \uplus W$, where $\uplus$ denotes
disjoint set union. Given a black and white graph $G=(B \uplus
W,E)$ and an integer $k$, the problem is to find a set of at most
$k$ vertices that dominate the black vertices. More formally, we
ask whether there is a subset $U \subseteq B \uplus W$, such that
$|U| \leq k$ and every vertex $v \in B-U$ satisfies $N(v) \cap U
\neq \emptyset$. Finally we give a new definition, specific to this
paper, for what it means to be a \textit{reduced} black and white
graph.

\begin{definition}
A black and white graph $G=(B \uplus W,E)$ is called reduced if it
satisfies the following conditions:
\begin{itemize}
\item $W$ is an independent set.

\item All the vertices of $W$ have degree at least $2$.

\item $N(w_1) \neq N(w_2)$ for every two distinct vertices
$w_1,w_2 \in W$.
\end{itemize}
\end{definition}

\section{Algorithms for the Dominating Set Problem}

\subsection{Degenerated Graphs}

The algorithm for degenerated graphs is based on the following
combinatorial lemma.

\begin{lemma}\label{combdeg}
Let $G=(B \uplus W,E)$ be a $d$-degenerated black and white graph.
If $|B| > (4d+2)k$, then there are at most $(4d+2)k$ vertices in
$G$ that dominate at least $|B|/k$ vertices of $B$.
\end{lemma}

\begin{proof}
Denote  $R = \{v \in B \cup W \big | |(N_G(v) \cup \{v\}) \cap B|
\geq |B|/k \}$. By contradiction, assume that $|R| > (4d+2)k$. The
induced subgraph $G[R \cup B]$ has at most $|R|+|B|$ vertices and
at least $\frac{|R|}{2} \cdot (\frac{|B|}{k}-1)$ edges. The
average degree of $G[R \cup B]$ is thus at least
$$
\frac{|R|(|B|-k)}{k(|R|+|B|)} \geq \frac{\min \{|R|,|B|\}}{2k} - 1
> 2d.
$$
This contradicts the fact that $G[R \cup B]$ is $d$-degenerated.
\qed 
\end{proof}

\begin{theorem}
There is a $k^{O(dk)} n$ time algorithm for finding a dominating
set of size at most $k$ in a $d$-degenerated black and white graph
with $n$ vertices that contains such a set.
\end{theorem}

\begin{proof}
The pseudocode of algorithm $DominatingSetDegenerated(G,k)$ that
solves this problem appears below. If there is indeed a dominating
set of size at most $k$, then this means that we can split $B$
into $k$ disjoint pieces (some of them can be empty), so that each
piece has a vertex that dominates it. If $|B| \leq (4d+2)k$, then
there are at most $k^{(4d+2)k}$ ways to divide the set $B$ into
$k$ disjoint pieces. For each such split, we can check in $O(kdn)$
time whether every piece is dominated by a vertex. If
$|B| > (4d+2)k$, then it follows from Lemma \ref{combdeg} that
$|R| \leq (4d+2)k$. This means that the search tree can grow to be
of size at most $(4d+2)^k k!$ before possibly reaching the
previous case. This gives the needed time bound.
\qed 
\end{proof}

\begin{algorithm}
\dontprintsemicolon \KwIn{Black and white $d$-degenerated graph $G=(B \uplus W,E)$,
integers $k,d$} \KwOut{A set dominating all vertices of $B$ of size at most $k$ or $NONE$
if no such set exists} \If{$B=\emptyset$}{\Return {$\emptyset$}}
\ElseIf{$k=0$}{ \Return{$NONE$} } \ElseIf{$|B| \leq (4d+2)k$}{
\ForAll{possible ways of splitting $B$ into $k$ (possibly empty)
disjoint pieces $B_1, \ldots ,B_k$}{ \If{each piece $B_i$ has a
vertex $v_i$ that dominates it}{ \Return{ $\{v_1, \ldots ,v_k\}$}}
} \Return{$NONE$}} \Else { $R \leftarrow \{v \in B \cup W \big |
|(N_G(v) \cup \{v\}) \cap B| \geq |B|/k \}$ \; \ForAll{$v \in R$}{
Create a new graph $G'$ from $G$ by marking all the elements of
$N_G(v)$ as white and removing $v$ from the graph \; $D \leftarrow
DominatingSetDegenerated(G',k-1)$ \; \If{$D \neq NONE$}{ \Return
{$D \cup \{v\}$} } } \Return {$NONE$}}
\caption{$DominatingSetDegenerated(G,k)$} \label{algdeg}
\end{algorithm}

\subsection{Graphs with an Excluded Minor}

Graphs with either an excluded minor or with no topological minor
are known to be degenerated. We will apply the following useful
propositions.

\begin{proposition} \label{degtop}
\cite{journals/ejc/BollobasT98,journals/cpc/KomloS96}
There exists a constant $c$ such that, for every $h$, every graph
that does not contain $K_h$ as a topological minor is $c
h^2$-degenerated.
\end{proposition}

\begin{proposition} \label{degmin}
\cite{journals/combinatorica/Kostochka84,MR735367,journals/jct/Thomason01}
There exists a constant $c$ such that, for every $h$, every graph
with no $K_h$ minor is $c h \sqrt{\log h}$-degenerated.
\end{proposition}

The following lemma gives an upper bound on the number of cliques
of a prescribed fixed size in a degenerated graph.

\begin{lemma}\label{clique}
If a graph $G$ with $n$ vertices is $d$-degenerated, then for
every $k \geq 1$, $G$ contains at most $\binom{d}{k-1} n$ copies
of $K_k$.
\end{lemma}

\begin{proof}
By induction on $n$. For $n=1$ this is obviously true. In the
general case, let $v$ be a vertex of degree at most $d$. The
number of copies of $K_k$ that contain $v$ is at most
$\binom{d}{k-1}$. By the induction hypothesis, the number of
copies of $K_k$ in $G-v$ is at most $\binom{d}{k-1} (n-1)$.
\qed 
\end{proof}

We can now prove our main combinatorial results.

\begin{theorem}\label{combtop}
There exists a constant $c>0$, such that for every reduced black
and white graph $G=(B \uplus W,E)$, if $G$ does not contain $K_h$
as a topological minor, then there exists a vertex $b \in B$ of
degree at most $(ch)^h$.
\end{theorem}

\begin{proof}
Denote $|B|=n>0$ and $d=c h^2$ where $c$ is the constant from
Proposition \ref{degtop}. Consider the vertices of $W$ in
some arbitrary order. For each such vertex $w \in W$, if there
exist two vertices $b_1,b_2 \in N(w)$, such that $b_1$ and $b_2$
are not connected, add the edge $\{b_1,b_2\}$ and remove
the vertex $w$ from the graph. Denote the resulting graph
$G'=(B \uplus W',E')$. Obviously, $G'[B]$ does not contain $K_h$
as a topological minor and therefore has at most $dn$ edges. The
number of edges in the induced subgraph $G'[B]$ is at least the
number of white vertices that were deleted from the graph, which
means that at most $dn$ were deleted so far.

We now bound $|W'|$, the number of white vertices in $G'$. It
follows from the definition of a reduced black and white graph that
there are no white vertices in $G'$ of degree smaller than $2$.
The graph $G'$ cannot contain a white vertex of degree $h-1$ or
more, since this would mean that the original graph $G$ contained
a subdivision of $K_h$. Now let $w$ be a white vertex of $G'$ of
degree $k$, where $2 \leq k \leq h-2$. The reason why $w$ was not
deleted during the process of generating $G'$
is because $N(w)$ is a clique of size $k$ in $G'[B]$.
The graph $G'$ is a reduced black and white graph, and therefore
$N(w_1) \neq N(w_2)$ for every two different white vertices $w_1$
and $w_2$. This means that the neighbors of each white vertex
induce a different clique in $G'[B]$. By applying Lemma
\ref{clique} to $G'[B]$, we get that the number of white vertices
of degree $k$ in $G'$ is at most $\binom{d}{k-1} n$. This means
that $|W'| \leq \left [\binom{d}{1} + \binom{d}{2} + \cdots +
\binom{d}{h-3} \right ] n$. We know that $|W| \leq |W'|+dn$ and
therefore $|E| \leq d(|B|+|W|) \leq d \left [3d + \binom{d}{2} +
\cdots + \binom{d}{h-3} \right ] n$. Obviously, there exists a
black vertex of degree at most $2|E|/n$. The result now follows by
plugging the value of $d$ and using the fact that $\binom{n}{k}
\le (\frac{en}{k})^k$.
\qed 
\end{proof}

\begin{theorem}\label{combmin}
There exists a constant $c>0$, such that for every reduced black
and white graph $G=(B \uplus W,E)$, if $G$ is $K_h$-minor-free,
then there exists a vertex $b \in B$ of degree at most $(c \log
h)^{h/2}$.
\end{theorem}

\begin{proof}
We proceed as in the proof of Theorem \ref{combtop} using
Proposition \ref{degmin} instead of Proposition \ref{degtop}.
\qed 
\end{proof}

\begin{theorem}\label{algtop}
There is an $(O(h))^{hk} n$ time algorithm for finding a
dominating set of size at most $k$ in a black and white graph with
$n$ vertices and no $K_h$ as a topological minor.
\end{theorem}

\begin{proof}
The pseudocode of algorithm $DominatingSetNoMinor(G,k)$ that
solves this problem appears below. Let the input be a black and
white graph $G=(B \uplus W,E)$. It is important to notice that the
algorithm removes vertices and edges in order to get a (nearly) reduced
black and white graph. 
This can be done in time $O(|E|)$ by a careful procedure based
on the proof of Theorem \ref{combtop} combined with radix sorting.
We omit the details which will appear in the full version of the paper.
\iffalse
We now show how this can be done in
$O(|E|)$ time. It is easy to ensure that every vertex $w \in W$
will have a sorted set of neighbors $N(w)$. The lists of neighbors
are now sorted using radix sort in linear time. Now we can
eliminate all duplicates, that is, ensure that $N(w_1) \neq
N(w_2)$ for every two different vertices $w_1,w_2 \in W$ that
remained in the graph. 
\fi
The time bound for the algorithm now
follows from Theorem \ref{combtop}.
\qed 
\end{proof}

\begin{algorithm}
\dontprintsemicolon \KwIn{Black and white ($K_h$-minor-free) graph $G=(B \uplus W,E)$,
integer $k$} \KwOut{A set dominating all vertices of $B$ of size at most $k$ or $NONE$
if no such set exists} \If{$B=\emptyset$}{ \Return{$\emptyset$}}
\ElseIf{$k=0$}{ \Return{$NONE$} } \Else { Remove all edges of $G$
whose two endpoints are in $W$ \; Remove all white vertices of $G$
of degree 0 or 1 \; As long as there are two different vertices
$w_1,w_2 \in W$ with $N(w_1)=N(w_2), ~|N(w_1)|<h-1$, remove one of them from the
graph \; Let $b \in B$ be a vertex of minimum degree among all
vertices in $B$ \; \ForAll{$v \in N_G(b) \cup \{b\}$} { Create a
new graph $G'$ from $G$ by marking all the elements of $N_G(v)$ as
white and removing $v$ from the graph \; $D \leftarrow
DominatingSetNoMinor(G',k-1)$ \; \If{$D \neq NONE$}{ \Return {$D
\cup \{v\}$}} } \Return {$NONE$} }
\caption{$DominatingSetNoMinor(G,k)$} \label{algminor}
\end{algorithm}

\begin{theorem}
There is an $(O(\log h))^{hk/2} n$ time algorithm for finding a
dominating set of size at most $k$ in a black and white graph with
$n$ vertices which is $K_h$-minor-free.
\end{theorem}

\begin{proof}
The proof is analogues to that of Theorem \ref{algtop} using
Theorem \ref{combmin} instead of Theorem \ref{combtop}.
\qed 
\end{proof}

\subsection{The Weighted Case}

In the weighted dominating set problem, each vertex of the graph
has some positive real weight. The goal is to find a dominating
set of size at most $k$, such that the sum of the weights of all
the vertices of the dominating set is as small as possible.
The algorithms we presented can be generalized to deal with the
weighted case without changing the time bounds. 
In this case, the whole search tree needs to be scanned and one cannot settle
for the first valid solution found.

Let $G=(B \uplus W,E)$ be the input graph to the algorithm. In
algorithm $\ref{algdeg}$ for degenerated graphs, we need to
address the case where $|B| \leq (4d+2)k$. In this case, the algorithm scans
all possible ways of splitting $B$ into $k$ disjoint pieces $B_1,
\ldots,B_k$, and it has to be modified, so that it will always
choose a vertex with minimum weight that dominates each piece. In
algorithm $\ref{algminor}$ for graphs with an excluded minor, the
criterion for removing white vertices from the graph is modified
so that whenever two vertices $w_1,w_2 \in W$ satisfy $N(w_1)=N(w_2)$, 
the vertex with the bigger weight is removed.

\section{Finding Induced Cycles}

\subsection{Degenerated Graphs}

Recall that $N_i^+(v)$ is the set of all vertices that can be
reached from $v$ by a directed path of length exactly $i$. If the
outdegree of every vertex in a directed graph $D=(V,A)$ is at most
$d$, then obviously $|N_i^+(v)| \leq d^i$ for every $v \in V$ and
$i \geq 1$.

\begin{theorem}
For every fixed $d \geq 1$ and $k \leq 5$, there is a
deterministic $O(n)$ time algorithm for finding an \textbf{induced} cycle
of length $k$ in a $d$-degenerated graph on $n$ vertices.
\end{theorem}

\begin{proof}
Given a $d$-degenerated graph $G=(V,E)$ with $n$ vertices, we
orient the edges so that the outdegree of all vertices is at most
$d$. This can be done in time $O(|E|)$. Denote the resulting
directed graph $D=(V,A)$. We can further assume that $V=\{1,2,
\ldots ,n\}$ and that every directed edge $\{u,v\} \in A$
satisfies $u < v$. This means that an out-neighbor of a vertex $u$
will always have an index which is bigger than that of $u$. We now
describe how to find cycles of size at most $5$.

To find cycles of size $3$ we simply check for each vertex $v$
whether $N^+(v) \cap N_2^+(v) \neq \emptyset$. Suppose now that we
want to find a cycle $v_1-v_2-v_3-v_4-v_1$ of size $4$. Without
loss of generality, assume that $v_1 < v_2 < v_4$. We distinguish
between two possible cases.
\begin{itemize}
\item $v_1 < v_3 < v_2 < v_4$: Keep two counters $C_1$ and $C_2$
for each pair of vertices. For every vertex $v \in V$ and every
unordered pair of distinct vertices $u,w \in N^+(v)$, such that
$u$ and $w$ are not connected, we raise the counter $C_1(\{u,w\})$
by one. In addition to that, for every vertex $x \in N^+(v)$ such
that $u,w \in N^+(x)$, the counter $C_2(\{u,w\})$ is incremented.
After completing this process, we check whether there are two
vertices for which $\binom{C_1(\{u,w\})}{2}-C_2(\{u,w\})
> 0$. This would imply that an induced $4$-cycle was found.
\item $v_1 < v_2 < v_3 < v_4$ or $v_1 < v_2 < v_4 < v_3$: Check
for each vertex $v$ whether the set $\{v\} \cup N^+(v) \cup
N_2^+(v) \cup N_3^+(v)$ contains an induced cycle.
\end{itemize}

To find an induced cycle of size $5$, a more detailed case
analysis is needed. It is easy to verify that such a cycle
has one of the following two types.
\begin{itemize}
\item There is a vertex $v$ such that $\{v\} \cup N^+(v) \cup
N_2^+(v) \cup N_3^+(v) \cup N_4^+(v)$ contains the induced cycle.
\item The cycle is of the form $v - x - u - y - w - v$, where $x
\in N^+(v)$, $u \in N^+(x) \cap N^+(y)$, and $w \in N^+(v) \cap
N^+(y)$. The induced cycle can be found by defining counters in a
similar way to what was done before. We omit the details.
\end{itemize}
\qed 
\end{proof}

The following simple lemma shows that a linear time algorithm for finding
an induced $C_6$ in a $2$-degenerated graph would imply that a
triangle (a $C_3$) can be found in a general graph in $O(|V|+|E|) \leq O(|V|^2)$
time. It is a long standing open question to improve the natural
$O(|V|^{\omega})$ time algorithm for this problem
\cite{SICOMP::ItaiR1978}.

\begin{lemma}
Given a linear time algorithm 
for finding an induced $C_6$
in a $2$-degenerated graph, it is possible to find triangles in
general graphs in $O(|V|+|E|)$ time.
\end{lemma}

\begin{proof}
Given a graph $G=(V,E)$, subdivide all the edges. The new graph obtained
$G'$ is $2$-degenerated and has $|V|+|E|$ vertices. A
linear time algorithm for finding an induced $C_6$ in $G'$
actually finds a triangle in $G$. By assumption, the running
time is $O(|V|+|E|) \leq O(|V|^2)$.
\qed 
\end{proof}

\subsection{Minor-Closed Families of Graphs}

\begin{theorem}\label{minor-rand}
Suppose that $G$ is a graph with $n$ vertices taken from some
non-trivial minor-closed family of graphs. For every fixed $k$, if
$G$ contains an \textbf{induced} cycle of size $k$, then it can be found in $O(n)$
expected time.
\end{theorem}

\begin{proof}
There is some absolute constant $d$, so that $G$ is
$d$-degenerated. Orient the edges so that the maximum outdegree is
at most $d$ and denote the resulting graph $D=(V,E)$. We now use
the technique of random separation. Each vertex $v \in V$ of the
graph is independently removed with probability $1/2$, to get some
new directed graph $D'$. Now examine some (undirected) induced
cycle of size $k$ in the original directed graph $D$, and denote
its vertices by $U$. The probability that all the vertices in $U$
remained in the graph and all vertices in $N^+(U)-U$ were removed from
the graph is at least $2^{-k(d+1)}$.

We employ the color-coding method to the graph $D'$. Choose a
random coloring of the vertices of $D'$ with the $k$ colors
$\{1,2, \ldots ,k\}$. For each vertex $v$ colored $i$, if $N^+(v)$
contains a vertex with a color which is neither $i-1$ nor $i+1
\pmod{k}$, then it is removed from the graph. For each induced cycle
of size $k$, its vertices will receive distinct colors and it
will remain in the graph with probability at least $2k^{1-k}$.

We now use the $O(n)$ time algorithm from \cite{JACM::AlonYZ1995}
to find a multicolored cycle of length $k$ in the resulting graph. If such a
cycle exists, then it must be an induced cycle. Since $k$ and $d$
are constants, the algorithm succeeds with some small constant
probability and the expected running time is as needed.
\qed 
\end{proof}

The next theorem shows how to derandomize this algorithm while
incurring a loss of only $O(\log n)$.

\begin{theorem}
Suppose that $G$ is a graph with $n$ vertices taken from some
non-trivial minor-closed family of graphs. For every fixed $k$,
there is an $O(n \log n)$ time deterministic algorithm for finding 
an \textbf{induced} cycle of size $k$ in $G$.
\end{theorem}

\begin{proof}
Denote $G=(V,E)$ and assume that $G$ is $d$-degenerated. We
derandomize the algorithm in Theorem \ref{minor-rand} using an
$(n,dk+k)$-family of perfect hash functions. 
This is a family of functions from $[n]$ to $[dk+k]$ such that for
every $S \subseteq [n]$, $|S|=dk+k$, there exists a function in the family
that is 1-1 on $S$.
Such a family of size
$e^{dk+k} (dk+k)^{O(\log (dk+k))} \log n$ can be efficiently
constructed \cite{FOCS::NaorSS1995}. We think of each function as
a coloring of the vertices with the $dk+k$ colors $C=\{1,2, \ldots
,dk+k\}$. For every combination of a coloring, a subset $L \subseteq C$ of $k$
colors and a bijection $f: L \to \{1,2, \ldots ,k\}$ the following
is performed. All the vertices that got a color from $c \in L$ now
get the color $f(c)$. The other vertices are removed from the
graph. 

The vertices of the resulting graph are colored with the $k$ colors
$\{1,2, \ldots ,k\}$. Examine some induced cycle of size $k$ in the
original graph, and denote its vertices by $U$. 
There exists some coloring $c$ in the family of perfect hash functions
for which all the vertices in $U \cup N^+(U)$ received different colors. 
Now let $L$ be the $k$ colors of the vertices in the cycle $U$ and let
$f: L \to [k]$ be the bijection that gives consecutive colors to vertices
along the cycle. 
This means that for this choice of $c$, $L$, and $f$, the induced cycle $U$
will remain in the graph as a multicolored cycle, whereas all the vertices 
in $N^+(U)-U$ will be removed from the graph.

We proceed as in the previous algorithm. Better dependence
on the parameters $d$ and $k$ can be obtained using the results in
\cite{journals/jct/AlonCKL03}.
\qed 
\end{proof}

\section{Concluding Remarks}\label{sec:conclude}

\begin{itemize}
\item The algorithm for finding a dominating set in graphs with an
excluded minor, presented in this paper, generalizes and improves known
algorithms for planar graphs and graphs with bounded genus. We
believe that similar techniques may be useful in improving and
simplifying other known fixed-parameter algorithms for graphs with an
excluded minor.
\item An interesting open problem is to decide whether there is a
$2^{O(\sqrt{k})} n^c$ time algorithm for finding a dominating set
of size $k$ in graphs with $n$ vertices and an excluded minor, where $c$ is some
absolute constant that does not depend on the excluded graph.
Maybe even a $2^{O(\sqrt{k})} n$ time algorithm can be achieved.
\end{itemize}

%\begin{thebibliography}{10}
%\bibliographystyle{splncs}
%\bibliographystyle{plain}
%\bibliography{shai}
%\end{thebibliography}

\end{document}